\documentstyle[epsf,aps,preprint]{revtex}
\begin{document}
\draft

\title{Two-dimensional Navier--Stokes simulation of
deformation and break up of liquid patches}

\author{St{\'e}phane Zaleski}
\author{Jie Li}
\author{Sauro Succi\cite{Succi_address}}
\address{Laboratoire de Mod\'elisation en M\'ecanique,\\
  CNRS URA 229,
Universit\'e Pierre et Marie Curie (Paris 6),\\
 Paris, France}

\date{February 22, 1995}
\maketitle

\begin{abstract}

The large deformations and break up of circular 2D liquid patches in a
high Reynolds number
($Re=1000$) gas flow are investigated numerically. The
2D, plane flow Navier--Stokes equations are directly solved with explicit tracking of the
interface between the two phases and a new algorithm for surface tension.
The numerical method is able to pursue the simulation beyond the breaking or
coalescence of droplets. The simulations are able
to unveil the intriguing details of the non-linear interplay between
the deforming droplets and the vortical structures in the droplet's wake.
\end{abstract}

The dynamical processes leading to breakup of a single lump of fluid
into several pieces and, conversely, the coalescence of several
pieces into a single one, have captured considerable attention
from both the theoretical \cite{Tannever94,Goldstein93,Constantin93}
and experimental side \cite{Stone94,Shi94,Brenner94}.
{}From the theoretical point of view, much of the interest stems from the
fact that breakup/coalescence provide eminent examples of topological
singularity formation in hydrodynamic systems.
Practical interest is even more evident if one thinks of the
enormous wealth of physico-chemical phenomena in which
drop formation and break up play a crucial role.
The merging of galaxies, spray vaporization
combustion in diesel engines and deformation of biological cells
are just but a few representative
examples.
In many instances the primary question is just to assess
under which conditions breakup/coalescence do occur.
However it is clear that
the  exploration of what happens after breakup/coalescence
is of paramount importance to deepen our understanding.
The latter question is even more formidable, but
the numerical simulations presented in this Letter should offer
a preliminary answer.

The physics of drop deformation and break up is governed by the
competition between hydrodynamic stresses, viscous or inertial, which act
to deform the droplet, and surface tension which opposes increase of
the surface
area, and thus tends to restore weakly deformed objects to a spherical shape.

In this article we present simulations of droplets of liquid moving about in
a gas environment. The mathematical idealisation of this problem is that of
a 2D, incompressible Newtonian flow with surface tension on the interface and viscous
dissipation in the bulk. The momentum balance equations are the Navier--Stokes
equations for an incompressible fluid of variable viscosity
\begin{equation}
\rho( \partial_t {\bf u} + {\hbox{\bf u}} \cdot \nabla {\hbox{\bf u}})
= - \nabla p
+ \nabla \cdot \left( \mu {\bf E} \right) +
 \nabla  \cdot \left[  \sigma ( {\bf I} - {\bf n} \otimes {\bf n})
\delta_S \right]
\label{NSE}
\end{equation}
where $E_{ij} = \partial_i u_j + \partial_j u_i$ is the rate of strain
tensor, $I_{ij} = \delta_{ij}$  the unit tensor,
$\delta_S$ a distribution concentrated on the interface,
${\bf n}$ is the normal to the interface,
and $\sigma$ the surface tension coefficient.
The fluid density $\rho$ and the
viscosity $\mu$  are constant in each phase but vary from phase to phase.
The specific form of the surface
tension term used is equivalent to the more classical Laplace law
\cite{Batchelor,Lafaurie94}.
We consider incompressible flow with $\nabla \cdot {\hbox{\bf u}} = 0$ , and
the interface moves at a normal velocity $U_I = {\hbox{\bf u}} \cdot {\bf n}$.
In
addition to these equations, a condition is needed for the reconnection of
interfaces. In a real flow, this reconnection is a complex process, involving
long range molecular interactions between interfaces.  It is
impossible with current computing capabilities to simulate both the large
scale,
high Reynolds flow around a droplet and the molecular interactions.
The pragmatic alternative is to introduce a cutoff scale
$\delta_C$ below which  one will not attempt to model the interface physics.
Such a cutoff is consistent with the spontaneous behavior of
the Volume of Fluid methods described below,
in which liquid sheets of thickness smaller than the mesh size $h$
tend to break.

Taking as a length scale the diameter $D$ of the droplet and a characteristic
speed $U$ of the flow, the problem
has 4 dimensionless numbers, the gas and liquid  Reynolds numbers
$Re_i = \rho_i U D / \mu_i$, $i=L,G$,
the gas Weber number ${\hbox{\it We}}_G = \rho_G U^2 D / \sigma$ and the
density ratio $\rho_L/ \rho_G$.
Experiments show that droplets suddenly placed in a high speed flow break
when ${\hbox{\it We}}_G$ is between $10$ and $20$
\cite{wierzba90,krzeczkowski80}.
Several droplet break up regimes have been identified
\cite{simpkins71,pilch87}, and the basic theory
involves various instability mechanisms for the liquid gas surface,
\cite{kitscha89,liu93}  following the pioneering work of Taylor
\cite{Taylor63}.
The authors are not aware, at this date, of numerical simulations of this
problem beyond relatively small droplet deformations. However it must
be noted that simulations exist in the limit of vanishing $Re_i$
\cite{Stone94}.

We used the method described in \cite{Lafaurie94,Li95}.
A first order in time explicit integration of equations
(\ref{NSE}) was performed using the
MAC staggered finite difference grid
for the momentum balance equation. The 2D version of the method was used in order
to achieve calculations on larger grids.
The incompressibility
condition is accurately met by a projection method
\cite{Chorin68} with the help of a multigrid algorithm \cite{Press91}.
Surface tension is implemented
in a momentum conserving way, via the introduction of a non-isotropic
stress tensor concentrated near the interface\cite{Lafaurie94}.
This representation
of surface tension stresses is especially interesting for
the simulation of break up, since it
avoids the singularity which would occur in the continuum limit
when interfaces change topology and the curvature becomes locally infinite.

\begin{figure}[p]
\epsfxsize= 9.5 cm
\centerline{\epsfbox{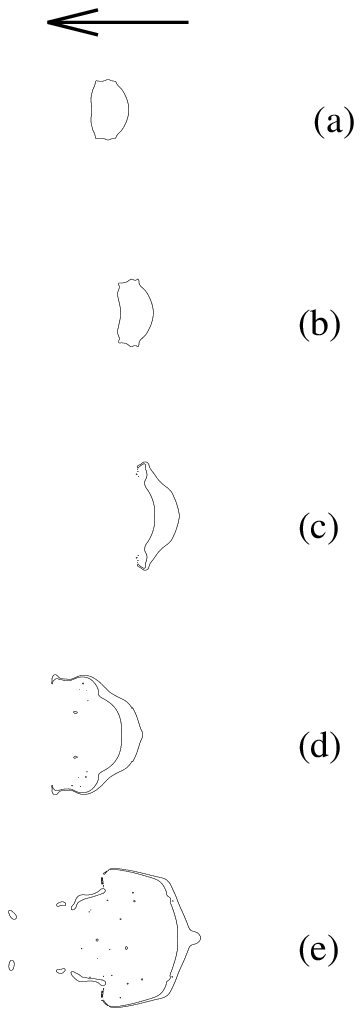}}
{\centerline{\parbox{15 cm}{\caption{{\protect \small Simulations of droplet
break up for
${\hbox{\it We}}_G=100$.
\label{h1fig}}}}}}
\end{figure}

The velocity field ${{\hbox{\bf u}}}$ obtained at each time step
 is used to propagate the interface using the Volume of Fluid method:
the location of the interface is
represented by the volume fraction $C_{ij}$ of fluid 1 in the computational
cell $i,j$
\cite{Youngs82,Puckett92,Poo91,Li95}
We have $0 < C_{ij} < 1$ in cells cut by the interface and
$C_{ij} = 0$ or $1$ away from it.
The propagation of the interface at velocity $U$ is performed
in several steps. In a first step, the interface is reconstructed in
each cell independently.
 Linear segments of slope approaching that of the interface
are constructed, in
the so-called Piecewise Linear Interface Construction (PLIC).
The construction uses a local interface normal ${\bf n}$
estimated using an 8 point (in 2D) centered
finite difference of $C_{ij}$.
In a second step, the interface motion is calculated in a Lagrangian
manner with velocities obtained by linear interpolation.
Finally the volume fractions are recalculated.
With the PLIC method\cite{Youngs82,Puckett92,Poo91,Li95},
the position of the interface is
reconstructed with errors of order ${\cal O}(\kappa\, h^2)$,
where $\kappa$ is the local curvature, for the position
of the interfaces, and thus more accurately than in most volume fraction
methods, including that of ref. \cite{Lafaurie94}.
During the simulations, we observe that only a
very small fraction of the mass is lost. 
In the complex case of Fig.~\ref{h1fig} this  fraction is
less than $2 10^{-3}$ over the entire simulation. 

\begin{figure}[p]
\epsfxsize= 9.5 cm
\centerline{\epsfbox{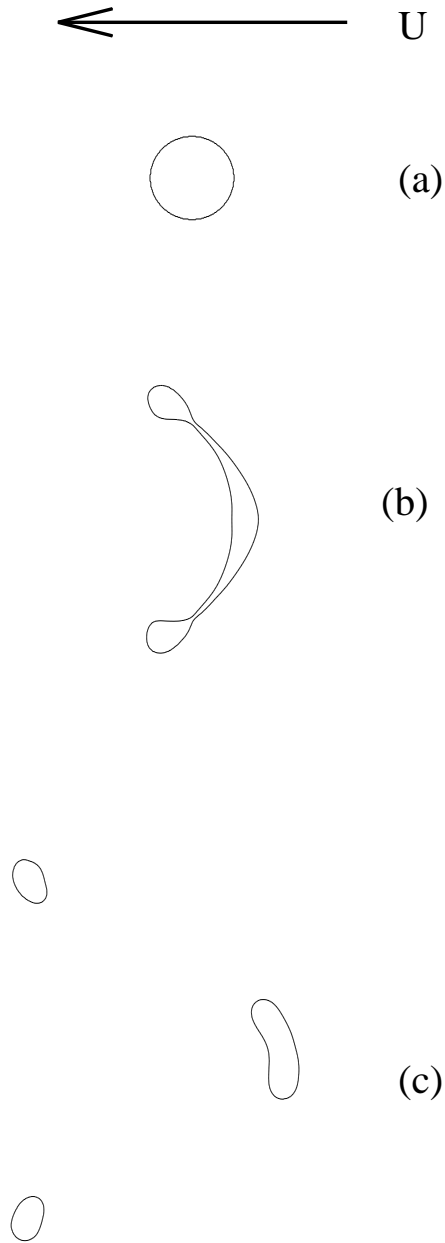}}
{\centerline{\parbox{15 cm}{\caption{{\protect \small Simulations of droplet
break up for ${\hbox{\it We}}_G=10$. $512^2$ grid points in a periodix box
are used.
\label{h2fig}}}}}} \end{figure}

\begin{figure}[p]
\epsfxsize= 15 cm
\centerline{\epsfbox{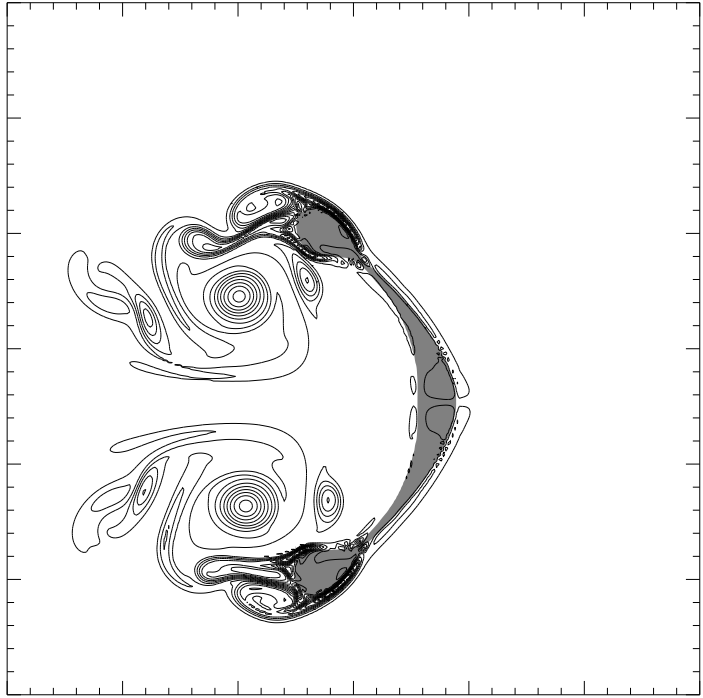}}
{\centerline{\parbox{15 cm}{\caption{{\protect \small Same as figure 1b,
but with vorticity
contours shown.
\label{h2velofig}}}}}}
\end{figure}

We  performed our simulations in a square periodic box of size $5D \times 5D$.
The simulation was initialized with a uniform velocity $U$ in the gas
and the liquid droplet at rest.
In all simulations reported here
we kept $Re_L=2,000$, the liquid to gas density ratio $\rho_L/\rho_G =10$ and
$Re_G=1,000$.
Several droplet deformation and break up scenarios
have been unveiled for varying ${\hbox{\it We}}_G$.
The first scenario is shown on
Fig.~\ref{h2fig}. As a result of the presence of two rear-vortices
engendered by the droplet motion, a concavity
develops in the droplet surface which takes a typical ``banana-shaped"
configuration, with its concave side facing downstream
(Fig~\ref{h2fig}(b) and Fig. \ref{h2velofig}).
These vortices further stretch the droplet until rupture occurs near the
tips. The coherent structures in the wake of the droplet
are characteristic of 2D turbulence. A boundary layer develops on the front of
the droplet as predicted by Taylor \cite{Taylor63}.
However at this ${\hbox{\it We}}_G$ the boundary layer is stable.

The ``mother'' droplet may again break in a similar fashion thus
generating additional children droplets.
Alternatively, if ${\hbox{\it We}}_G$ is sufficiently high second-generation
droplets may also break up, producing third-generation droplets
 in a kind of bifurcation cascade.
Moderate resolution ($256^2$) and high resolution ($512^2$)
 simulations exhibited similar results.

Simulations at larger Weber numbers such as the ${\hbox{\it We}}_G = 100$
simulation of Fig. \ref{h1fig} show the formation
of much smaller scale structures.
The droplet forms elongated filaments. The boundary layer on the front side
of the droplet is now unstable, and  horn--like structures typical of the
Kelvin--Helmholtz instability are seen to grow while they are transported
downstream  (see Fig.~\ref{h1fig}(b)).

\begin{figure}[p]
\epsfxsize= 7 cm
\centerline{\epsfbox{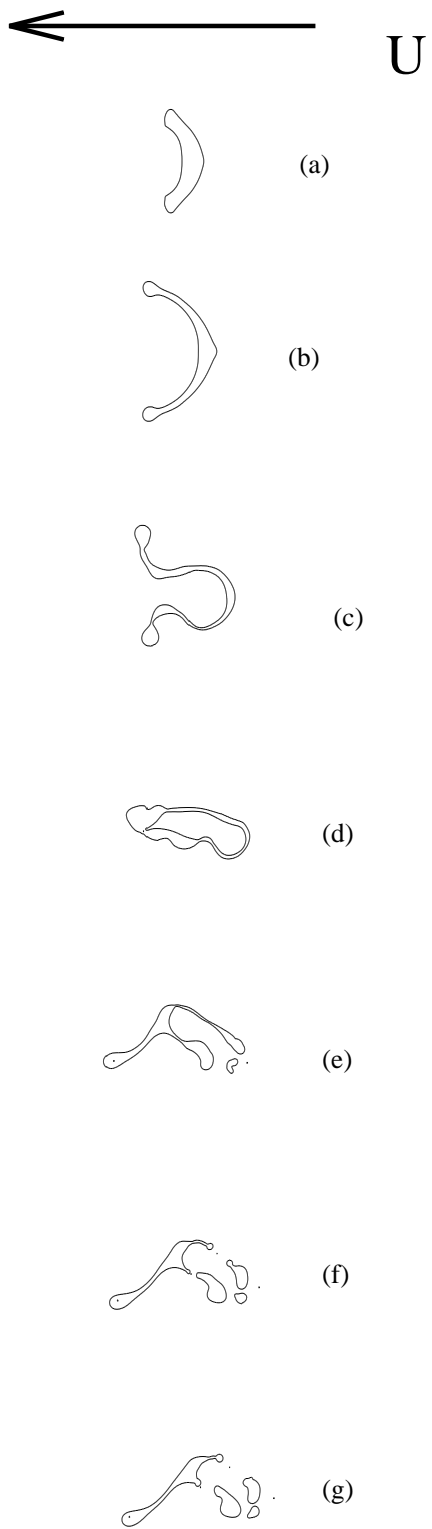}}
{\centerline{\parbox{15 cm}{\caption{{\protect \small Simulations of droplet
break up for
${\hbox{\it We}}_G=20$. \label{h3fig}}}}}}
\end{figure}

Yet another mechanism is shown on Fig.~\ref{h3fig}. There the
droplet elongates then makes a bag with its concave side facing downstream.
The bag closes, then breaks on the upstream side before
yielding several separated droplets.

The simulation, although 2D, may be qualitatively compared to the experimental
results. There are qualitative similarities, such as the breaking
near the tips on Fig.~\ref{h2fig}(b) or the bag formation.
The ${\hbox{\it We}}_G=100$ simulation shows the sheet stripping mode
mechanism reported by \cite{pilch87} for $100 < {\hbox{\it We}}_G < 350$.
The most important difference occurs at ${\hbox{\it We}}_G=10$ where
in experiments the concave side faces mostly upstream.
We believe that this difference arises because of the initial
conditions we use, which result in a jump of the velocity
--- a vortex sheet --- at the interface.
This vortex sheet rolls up behind the droplet and creates structure seen on
Fig. \ref{h2velofig}).
Work is in progress to investigate the influence of initial conditions.


In conclusion, the results highlight the power of
advanced numerical techniques
to unveil the fascinating complexity  resulting from
the nonlinear interplay between gas-liquid interface and gas vortex motion.
In particular, the crucial role played by coherent vortical structures
suggests that the inclusion of the $Re_G$  dependence is key
to the formulation of more advanced and realistic break up criteria.
On the other hand, these results also indicate that
caution is needed before the results provided by the numerical tool can be
effectively converted into quantitative information of engineering
interest, such as phase-diagrams and similar data.

One difficulty rests with the slow convergence of such calculations
with the number of grid-points.
In some regimes, the lower resolution experiments
produce a similar picture as provided by higher--resolution ones, the
main difference being that transitions
in phase--diagrams occur at different ${\hbox{\it We}}_G$.
In some instances, however, genuinely new mechanisms arise: the
bag mechanism of Fig.~\ref{h3fig} was observed only in $512^2$ simulations.
A second difficulty is the two-dimensional nature of these calculations.
As is well known, coherent vortical structures behave quite differently
in two with respect to three-dimensions, and so should break up mechanisms.
The Rayleigh Instability, 
which plays such an important role in the final stages
of capillary driven break up\cite{Shi94}, is absent in 2D.
(However this may affect only the smallest scales of a large $We$ break up).
Despite these difficulties,
we regard two-dimensional simulations as a very useful warm-up
for more realistic three-dimensional investigations.
While the former are already rather computationally expensive
(about 3 CPU seconds per time step for a $512^2$ resolution
on a IBM RS/6000 mod. 590 workstation), the latter set a pressing demand for
high-resolution computations requiring the use of the most powerful
present-day supercomputers.

\section*{Acknowledgements}

DRET, CNES and SEP are gratefully acknowledged for their funding.
The authors wish to thank Daniel Lhuillier for illuminating discussions.
S. Succi wishes to thank LMM for their kind hospitality and financial support.

\bibliographystyle{prsty}

\end{document}